\documentclass[letterpaper, 10 pt, conference]{ieeeconf}

\IEEEoverridecommandlockouts
\usepackage{graphics}
\usepackage{epsfig}
\usepackage{mathptmx}
\usepackage{times}
\usepackage{amsmath}
\usepackage{amssymb}

\title{\LARGE \bf
Control Energy of Lattice Graphs
}

\author{Isaac Klickstein and Francesco Sorrentino%
\thanks{I. Klickstein is a PhD student in the Department of Mechanical Engineering, University of New Mexico, Albuquerque, NM 87131, USA {\tt\small iklick@unm.edu}}%
\thanks{F. Sorrentino is with Faculty Member of the Department of Mechanical Engineering, University of New Mexico, Albuquerque, NM 87131, USA {\tt\small fsorrent@unm.edu}}%
\thanks{Both authors are supported by the National Science Foundation through NSF grant CMMI- 1400193, NSF grant CRISP- 1541148 and ONR Award No. N00014-16-1-2637.}
}

\begin{document}
  \maketitle
  \thispagestyle{empty}
  \pagestyle{empty}
  \begin{abstract}
    The control of complex networks has generated a lot of interest in a variety of fields from traffic management to neural systems.
    A commonly used metric to compare two particular control strategies that accomplish the same task is the control energy, the integral of the sum of squares of all control inputs.
    The minimum control energy problem determines the control input that lower bounds all other control inputs with respect to their control energies.
    Here, we focus on the infinite lattice graph with linear dynamics and analytically derive the expression for the minimum control energy in terms of the modified Bessel function.
    We then demonstrate that the control energy of the infinite lattice graph accurately predicts the control energy of finite lattice graphs.
  \end{abstract}
  \section{INTRODUCTION}
  The control of complex networks has remained an active area of research \cite{shirin2017optimal,chen2016energy,ruths2014control,liu2016control,pasqualetti2014controllability}.
  Applications are found in diverse fields from power grids \cite{menck2014dead} to marketing on social networks \cite{proskurnikov2016opinion} to networked autonomous vehicles \cite{olfati2007consensus} and many others.
  Recent results have shown that the control energy required for complex networks scales exponentially with respect to both the number of driver nodes and/or the number of target nodes \cite{klickstein2017energy, iudice2015structural, yan2015spectrum}.
  While there are many numerical experiments demonstrating the scaling behavior \cite{klickstein2017energy,yan2015spectrum,yan2012controlling}, and some attempts to explain the precise scaling exponents for ensembles of canonical model graphs, there is very little work attempting to derive analytical expressions for the control energy.
    Instead, a number of heuristics and greedy approximation algorithms to optimally place control inputs in networked systems have been developed recently \cite{cortesi2014submodularity,summers2016submodularity,olshevsky2014minimal,tzoumas2016minimal}.\\
  \indent
  Here we first consider lattice networks which have varied applications depending on the dimension of the lattice.
  One dimensional lattices are used to model serial processes where each nodal system depends only on the previous and subsequent systems.
  Two dimensional lattices are used to model dense planar systems such as transmission lines in a power grid or the road network in a city.
  Two and three dimensional lattices are used when one creates a mesh to solve PDEs such as the Laplace equation or the Poisson equation where our result would apply to a time-varying, controllable, source term.\\
  \indent
  We create two subsets of the nodes in these networks.
  The set of driver nodes consists of those nodes which receive an external control input directly.
  The set of target nodes have an assigned state at a final time when the control action ends.\\
  \indent
  In this paper, we derive the analytic, time-varying values of the controllability Gramian associated with networked dynamical systems on $d$-dimensional lattices.
  Also, for the case when the number of target nodes is small, we can write the analytic expression for the control energy and describe how the energy scales with the distance between driver nodes and target nodes.
  We demonstrate these results on a finite lattice graph.
  \section{BACKGROUND}
  \subsection{Graphs}
  A graph $\mathcal{G} = (\mathcal{V},\mathcal{E})$ is defined by a possibly infinite set of nodes $\mathcal{V}$ and set of edges $\mathcal{E} \subseteq (\mathcal{V} \times \mathcal{V})$ where if $(v_i,v_j) \in \mathcal{E}$ then node $v_i$ receives a signal from node $v_j$.
  In this work we assume that the graph is undirected, that is, if the edge $(v_i,v_j) \in \mathcal{E}$, then it implies that $(v_j,v_i) \in \mathcal{E}$ as well.
  Each edge also has a uniform edge weight of value $s > 0$.
  We also assume that each node has a weighted self-loop, $(v_i,v_i) \in \mathcal{E}$ for all $v_i \in \mathcal{V}$ of value $-p$ where $p > 0$.
  Also, for succinctness in the derivations, let $\mathcal{N}_i$ denote the set of neighbors of node $v_i$, that is, if $v_j \in \mathcal{N}_i$, then $(v_i,v_j) \in \mathcal{E}$.
  If the graph has a finite number of nodes, $n$, then the adjacency matrix $A$ can be defined as having element $A_{i,j} = s$ if $(v_i,v_j) \in \mathcal{E}$ and $A_{i,j} = 0$ otherwise when $ i \neq j$.
  All diagonal elements $A_{i,i} = -p$.
  \subsection{Minimum Energy Control of Linear Systems}
  Linear dynamical networks, where the dependencies of the evolution of the nodes' states can be described by the connectivity of a graph, are present in many different fields from the social sciences to physics and engineering.
  We include a set of control inputs $u_k(t)$, each connected to a single node, $\mathcal{D} \subseteq \mathcal{V}$, of size $|\mathcal{D}| = m$, denoted as the driver nodes.
  The time evolution $x_i(t)$ for the state of each node $v_i \in \mathcal{V}$ is defined as,
  \begin{equation}
    \dot{x}_i(t) = \sum_{j = 1}^n A_{i,j} x_j(t) + \sum_{v_k \in \mathcal{D}} \delta_{i,k} u_k(t)
  \end{equation}
  where $\delta_{i,k}$ is the Kronecker delta.
  Another  finite subset of the nodes, $\mathcal{T} \subseteq \mathcal{V}$, denoted as the target nodes, have an assigned desirable state at time $t_f$, denoted $x_{i,f}$.
  Our goal is to design a set of $m = |\mathcal{D}|$ control signals $u_k(t)$, $v_k \in \mathcal{D}$, such that at time $t_f$, $x_i(t_f) = x_{i,f}$ for all $v_i \in \mathcal{T}$.
  When the number of nodes, $|\mathcal{V}|$, is finite, we can write the optimal control problem that minimizes the control energy while driving the states of the nodes in the target set to some final state $x_i(t_f) = x_{i,f}$ for all $i \in \mathcal{T}$.
  Define the vector of all states with an assigned final value as $\textbf{y}(t) = C \textbf{x}(t)$ so that $C$ effectively selects those states associated with target nodes \cite{klickstein2017energy}.\\
  \indent
  The particular control input we are interested in is the solution to the minimum energy optimal control problem as it lower bounds all other control inputs in their $\mathcal{L}2$-norm that perform the same task (initial condition to final output).
  \begin{equation}\label{eq:optcon}
    \begin{aligned}
      \min && &J = \frac{1}{2} \int_0^{t_f} \sum_{k \in \mathcal{D}} u_k(t)^2 dt\\
      \text{s.t.} && &\dot{\textbf{x}}(t) = A \textbf{x}(t) + B \textbf{u}(t)\\
      && &x_i(0) = x_{i,0}\ \forall v_i \in \mathcal{V}\\
      && &y_i(t_f) = y_{i,f},\ \textbf{y}(t_f) = C \textbf{x}(t_f)
    \end{aligned}
  \end{equation}
  The matrix $A$ is the adjacency matrix of the graph and the $|\mathcal{V}| \times |\mathcal{D}|$ matrix $B$ denotes which control signals are attached to which driver nodes.
  The solution to the optimal control problem in Eq. \ref{eq:optcon} \cite{klickstein2017energy} is,
  \begin{equation}\label{eq:control}
    \textbf{u}(t) = B^T e^{A^T(t_f-t)} C^T \left(CW(t_f)C^T\right)^{-1} \left( \textbf{y}_f - C e^{A(t_f)} \textbf{x}_0\right)
  \end{equation}
  where $\textbf{y}_f$ contains all of the prescribed final states $x_{i,f}$ for all $v_i \in \mathcal{T}$, $\textbf{x}_0$ contains all of the prescribed initial conditions and $W(t_f)$ is the controllability Gramian which can be found by solving the differential Lyapunov equation,
  \begin{equation}\label{eq:fingram}
    \begin{aligned}
      \dot{W}(t) = AW(t) + W(t)A^T + BB^T, && W(0) = O_{n \times n}
    \end{aligned}
  \end{equation}
  The controllability Gramian is symmetric and positive semi-definite for $t > 0$.
  We assume that the triplet $(A,B,C)$ is output controllable (i.e., the rank of the matrix $[CB | CAB | \cdots |CA^{n-1}B]$ is equal to $|\mathcal{T}|$) which implies the matrix $CW(t_f)C^T$ is non-singular, and thus we may perform the inversion in Eq. \ref{eq:control}.
  The minimum control energy associated with the control inputs $\textbf{u}(t)$ in Eq. \ref{eq:control} can be found from the quadratic form,
  \begin{equation}\label{eq:energy}
    E = \int_{0}^{t_f} \textbf{u}^T(t) \textbf{u}(t) dt = \textbf{b}^T \left( CW(t_f) C^T \right)^{-1} \textbf{b}
  \end{equation}
  where the control action $\textbf{b} = \left( \textbf{y}_f - C e^{At_f} \textbf{x}_0\right)$ represents the difference between the desired final states and the final states if there were no control input for the nodes $v_i \in \mathcal{T}$.
  In general, for arbitrary adjacency matrix $A$ and set of driver nodes represented by $B$, the controllability Gramian cannot be computed analytically.
  The evolution of the individual elements of the controllability Gramian depends on both the topology of the graph and the distribution of control inputs so the energy of any control action, $E$, becomes extremely difficult to predict despite its importance in determining the required resources to perform the control action.
  \subsection{Modified Bessel Functions of the First Kind (MBFFK)}
  The results in the following sections are written in terms of the MBFFK, $I_{n}(z)$, of integer order $n$ \cite{abramowitz1967handbook}.
  \begin{equation}
    I_{n}(z) = \frac{1}{2\pi} \int_{-\pi}^{\pi} e^{-\mathcal{I} n \theta} e^{z \cos \theta} d\theta
  \end{equation}
  where $\mathcal{I} = \sqrt{-1}$ is the imaginary unit.
  There are a number of methods to compute $I_n(z)$ depending on the magnitude of $n$ and $z$ \cite{amos1986algorithm} such as by its series expansion \cite{abramowitz1967handbook} and many libraries exist which can compute the MBFFK such as the Gnu Scientific Library \cite{gough2009gnu}.  
  Some important properties of the MBFFK include,
  \begin{enumerate}
    \item $I_{-n}(z) = I_{n}(z)$ for any integer $n$ and complex argument $z$.
    This property preserves the symmetry visible in a lattice graph.
    \item $I_{n}(z) \geq I_{m}(z)$ for $n < m$ and real argument $z > 0$, so that as the order of the MBFFK increases, its value decreases.
    \item $\frac{\partial}{\partial z} I_n(z) > 0$ for any integer order $n$ and real argument $z > 0$.
    This implies the MBFFK is a strictly increasing function for $z > 0$.
  \end{enumerate}
  In the following sections, we first compute analytically the controllability Gramian for the infinite $2$-dimensional lattice graph in terms of an integral of a product of MBFFKs.
  We then generalize the derivation to any $d$-dimensional lattice.
  Finally, we apply our results to estimate the control energy for a finite lattice.
  \section{Results}
  \begin{figure}
    \centering
    \includegraphics[width=\columnwidth]{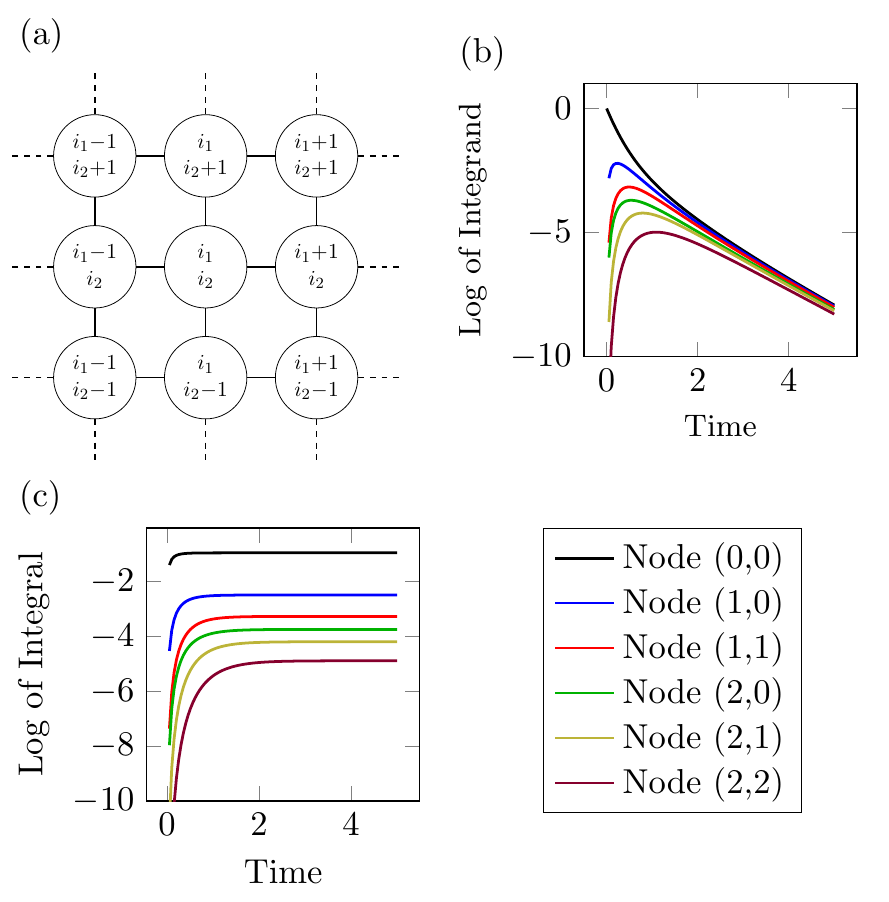}
    \caption{(a) An example of the indexing scheme we use to label the nodes in a lattice. For a two-dimensional lattice, a node is labeled as $\textbf{i} = (i_1,i_2)$.
    The eight nodes around node $\textbf{i}$ are shown with their labels as well.
    (b) For parameters $p = 5$ and $s = 1$, several time trajectories of the integrand in Eq. \ref{eq:2donedriver} for diagonal elements of the controllability Gramian.
    When $p > 2ds$, we see the typical behavior where the integrand reaches some maximum value before exponentially approaching zero.
    (c) For the same parameters, several time trajectories of the integral in Eq. \ref{eq:2donedriver} for the same diagonal elements of the controllability Gramian.
    When $p > 2ds$, the controllability Gramian equation is stable, that is, there exists a fixed point, which is clearly seen for $p = 5$ and $d = 2$.
    }
    \label{fig:cell}
  \end{figure}
  We now focus on graphs with an infinite number of nodes.
  As such, our notion of an adjacency matrix must be modified, and thus our definition of the controllability Gramian.
  By examining the individual entries of the controllability Gramian in Eq. \ref{eq:fingram}, we can express them in terms of the neighbors of each node, $\mathcal{N}_i$.
  \begin{equation}\label{eq:infgram}
    \begin{aligned}
    \dot{W}_{i,j} &= -2p W_{i,j} + \sum_{k \in \mathcal{N}_i} s W_{k,j} + \sum_{k \in \mathcal{N}_j} s W_{i,k} + \sum_{k \in \mathcal{D}} \delta_{i,k} \delta_{j,k}\\
      &-\infty < i_1,i_2,j_1,j_2 < \infty
    \end{aligned}
  \end{equation}
  where $s$ is the weight of each edge in the graph and the initial condition $W_{i,j} = 0$ for all nodal indices $-\infty < i, j < \infty$.
  While an analytical solution of Eq. \ref{eq:infgram} for a general graph does not exist, it can be computed for some graphs which have a regular connectivity pattern.
  It was recently shown that Eq. \ref{eq:infgram} for an infinite path graph can be solved analytically so that one can compute the energy as written in Eq. \ref{eq:energy} for an arbitrary control action.
  Here, we turn to $d$-dimensional lattice graphs (where the path graph can be thought of as a $1$-dimensional lattice).
  Of particular interest is the $2$-dimensional regular lattice which can be used to represent many planar systems such as road networks, infrastructure systems, printed circuitboards, cellular automata, and others.
  \subsection{Two-Dimensional Lattice}
  Before approaching the general $d$-dimensional lattice, we first derive in detail the controllability Gramian of the $2$-dimensional lattice.
  Each node in a $2$-dimensional lattice is connected to its four nearest neighbors as seen in Fig. \ref{fig:cell}(a).
  To label the nodes in a lattice, we employ a vector index $\textbf{i} = \left[ i_1\ i_2 \right]$, which in $2$-dimensions represents the node's coordinates from some reference node with index $\textbf{i}_{ref} = \left[ 0\ 0 \right]$.\\
  \indent
  In the rest of this section, we will generically use the notation $\textbf{j}$ to indicate the vector of indices $\textbf{j} = (j_1,j_2)$.
  Note that $j_1$ and $j_2$ may be negative integers as well.
  We introduce a mapping from the set of all integers $\mathbb{Z}$ to the pairs of integers $\mathbb{Z}^2$ (as both $\mathbb{Z}$ and $\mathbb{Z}^2$ are countably infinite).
  Then, even though the controllability Gramian as defined in Eq. \ref{eq:infgram} is a matrix, we are able to index is elements in the following derivation as $W_{\textbf{i},\textbf{j}} = W_{i_1,i_2,j_1,j_2}$, which is a tetradic and provides a mapping from each node pair $\textbf{i}$, $\textbf{j}$ to their lattice coordinates $i_1,i_2$ and $j_1,j_2$.
  Let us specialize Eq. \ref{eq:infgram} to the $2$-dimensional lattice by using the notation $\textbf{i}^{(k)} = \left[i_1+\delta_{1,k}\ i_2+\delta_{2,k}\right]$ and $\textbf{i}^{(-k)} = \left[i_1-\delta_{1,k}\ i_2-\delta_{2,k} \right]$ for $k \in \{1,2\}$ to represent the neighbors of node $\textbf{i}$.
  \begin{equation}\label{eq:2ddgram}
    \begin{aligned}
    \dot{W}_{\textbf{i},\textbf{j}} &= -2p W_{\textbf{i},\textbf{j}} + s \sum_{k=1}^2 \left( W_{\textbf{i}^{(k)},\textbf{j}} + W_{\textbf{i},\textbf{j}^{(k)}} + W_{\textbf{i}^{(-k)},\textbf{j}} + W_{\textbf{i},\textbf{j}^{(-k)}} \right)\\
    &+ \sum_{\textbf{k} \in \mathcal{D}} \delta_{\textbf{i},\textbf{k}} \delta_{\textbf{j},\textbf{k}}
    \end{aligned}
  \end{equation}
  The definition of the Kronecker delta is generalized to handle vector indices so that $\delta_{\textbf{i},\textbf{j}} = 1$ if both $i_1 = j_1$ and $i_2 = j_2$, and $\delta_{\textbf{i},\textbf{j}} = 0$ otherwise.
  Note that each derivative $\dot{W}_{\textbf{i},\textbf{j}} \equiv \dot{W}_{i_1,i_2,j_1,j_2}$ in Eq. \ref{eq:2ddgram} depends on both the current value of $W_{\textbf{i},\textbf{j}}$ as well as on its $4d=8$ neighbors.
  The differential equation in Eq. \ref{eq:2ddgram} is a linear nonhomogeneous equation which implies that we can solve the homogeneous equation first, and then use a convolution integral to account for the nonhomogeneous term.
  The derivation that continues from this point is for the homogeneous problem (which can also be thought of as the case when $\mathcal{D} = \emptyset$).
  The expression for the time derivative for each element $W_{\textbf{i},\textbf{j}}(t)$ of the controllability Gramian is decoupled from its neighbors by using the $4$-dimensional discrete time Fourier transform (DTFT) defined as,
  \begin{equation}\label{eq:2ddtft}
    \hat{W}_{\hat{\textbf{i}},\hat{\textbf{j}}}(t) = \sum_{\textbf{i},\textbf{j}} e^{\mathcal{I}i_1\hat{i}_1} e^{\mathcal{I}i_2\hat{i}_2} e^{\mathcal{I}j_1\hat{j}_1}e^{\mathcal{I}j_2\hat{j}_2} W_{\textbf{i},\textbf{j}}(t),
  \end{equation}
  where we use $\mathcal{I} = \sqrt{-1}$.
  \begin{figure*}
    \centering
    \includegraphics[width=\textwidth]{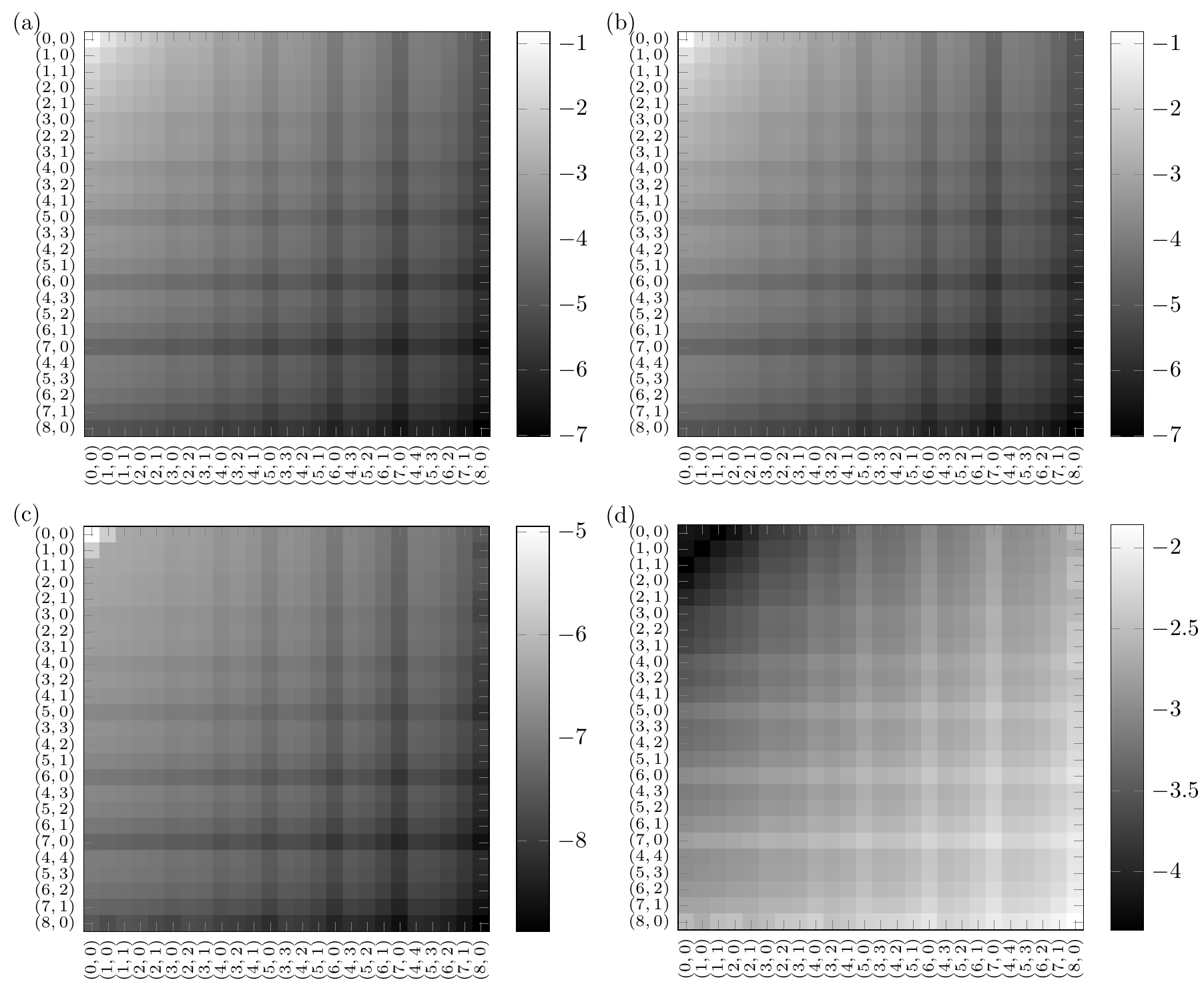}
    \caption{The controllability gramian of an infinite two dimensional lattice versus a finite two dimensional lattice of dimension $21 \times 21$.
    The parameters chosen are $s = 1$, $p = 5$, and $t = 5$.
    (a) The logarithm of the elements of the controllability gramian for a finite lattice.
    Note that each node is indexed by its position from the central node, which is the sole driver node.
    (b) The logarithm of the elements of the controllability gramian for an infinite lattice using Eq. \ref{eq:2donedriver}.
    While (a) and (b) appear qualitatively similar, the logarithm of the absolute error (c) is largest at element $W_{0,0,0,0}$.
    More importantly, the logarithm of the relative error (d) is smallest for element $W_{0,0,0,0}$ and grows larger as we choose nodes closer to the edge of the finite lattice.
    Please note the different log scales of panels (c) and (d).}
    \label{fig:error}
  \end{figure*}
  The transformed Gramian entries $\hat{W}_{\hat{\textbf{i}},\hat{\textbf{j}}}$ represent the modes of the original controllability Gramian.
  The notation $\hat{\textbf{i}}$ represents the indices of the transformed entries $\hat{W}_{\hat{\textbf{i}},\hat{\textbf{j}}}$ in order to differentiate them from the indices of the original entries, $\textbf{i}$.
  Applying the transformation in Eq. \ref{eq:2ddtft} to the dynamical system in Eq. \ref{eq:2ddgram} and simplifying, the decoupled differential equation becomes,
  \begin{equation}\label{eq:Whatij}
    \begin{aligned}
      \dot{\hat{W}}_{\hat{\textbf{i}},\hat{\textbf{j}}} &= \left( -2p + se^{-\mathcal{I} \hat{i}_1} + se^{\mathcal{I} \hat{i}_1} + se^{-\mathcal{I} \hat{i}_2} + se^{-\mathcal{I} \hat{i}_2}\right.\\
      &+ \left. se^{-\mathcal{I} \hat{j}_1} + se^{\mathcal{I} \hat{j}_1} + se^{-\mathcal{I} \hat{j}_2} + se^{\mathcal{I} \hat{j}_2} \right) \hat{W}_{\hat{\textbf{i}},\hat{\textbf{J}}}\\
      &= \left(-2p + 2s\cos \hat{i}_1 + 2s \cos \hat{i}_2 + 2s \cos \hat{j}_1 \right.\\
      &+ \left. 2s \cos \hat{j}_2 \right) \hat{W}_{\hat{\textbf{i}},\hat{\textbf{j}}}
    \end{aligned}
  \end{equation}
  As Eq. \ref{eq:Whatij} is simply a linear equation for $\hat{W}_{\hat{\textbf{i}},\hat{\textbf{j}}}$ decoupled from any other $\hat{W}_{\hat{\textbf{k}},\hat{\textbf{l}}}$, $\hat{\textbf{k}} \neq \hat{\textbf{i}}$ and $\hat{\textbf{j}} \neq \hat{\textbf{l}}$, each element $\hat{W}_{\hat{\textbf{i}},\hat{\textbf{j}}}$ can be solved for individually.
  \begin{equation}
    \hat{W}_{\hat{\textbf{i}},\hat{\textbf{j}}}(t) = e^{-2pt}e^{2st\cos \hat{i}_1} e^{2st \cos \hat{i}_2} e^{2st \cos \hat{j}_1} e^{2st \cos \hat{j}_2} \hat{W}_{\hat{\textbf{i}},\hat{\textbf{j}}}(0)
  \end{equation}
  With the solution in the $\hat{\textbf{i}},\hat{\textbf{j}}$ domain now known, we apply the inverse $4$-dimensional DTFT to find the solution in the $\textbf{i},\textbf{j}$ domain.
  \begin{equation}\label{eq:homo}
    \begin{aligned}
    W_{\textbf{i},\textbf{j}}(t) &= \frac{1}{(2\pi)^4} \int_{-\pi}^{\pi} \int_{-\pi}^{\pi} \int_{-\pi}^{\pi} \int_{-\pi}^{\pi} e^{-\mathcal{I} i_1 \hat{i}_1} e^{-\mathcal{I}i_2 \hat{i}_2} e^{-\mathcal{I} j_1 \hat{j}_1}\\ 
    &\times e^{-\mathcal{I} j_2 \hat{j}_2} e^{-2pt} e^{2st \cos \hat{i}_1} e^{2st \cos \hat{i}_2} e^{2st \cos \hat{j}_1} e^{2st \cos \hat{j}_2}\\
    &\times \sum_{\textbf{k},\textbf{l}} e^{\mathcal{I} \hat{i}_1 k_1} e^{\mathcal{I} \hat{i}_2 k_2} e^{\mathcal{I} \hat{j}_1 l_1} e^{\mathcal{I} \hat{j}_2 l_2} W_{\textbf{k},\textbf{l}}(0) d\hat{i}_1 d\hat{i}_2 d\hat{j}_1 d\hat{j}_2\\
    &= \sum_{\textbf{k},\textbf{l}} W_{\textbf{k},\textbf{l}}(0) e^{-2pt} \left( \frac{1}{2\pi} \int_{-\pi}^{\pi} e^{-\mathcal{I} \hat{i}_1 (i_1-k_1)} e^{2st \cos \hat{i}_1} d\hat{i}_1 \right)\\
    &\times \left( \frac{1}{2\pi} \int_{-\pi}^{\pi} e^{-\mathcal{I} \hat{i}_2 (i_2-k_2)} e^{2st \cos \hat{i}_2} d\hat{i}_2 \right)\\
    &\times \left( \frac{1}{2\pi} \int_{-\pi}^{\pi} e^{-\mathcal{I} \hat{j}_1 (j_1-l_1)} e^{2st \cos \hat{j}_1} d\hat{j}_1 \right)\\
    &\times \left( \frac{1}{2\pi} \int_{-\pi}^{\pi} e^{-\mathcal{I} \hat{j}_2 (j_2-l_2)} e^{2st \cos \hat{j}_2} d\hat{j}_2 \right)\\
    &= e^{-2pt} \sum_{\textbf{k},\textbf{l}} I_{i_1-k_1}(2ft) I_{i_2-k_2}(2st) I_{j_1-l_1}(2st)\\
    &\times I_{j_2-l_2}(2st) W_{\textbf{k},\textbf{l}}(0)
    \end{aligned}
  \end{equation}
  The solution is an infinite sum over all initial conditions scaled by the product of four MBFFKs whose order is the absolute distance between nodes $\textbf{i}$ and $\textbf{k}$ and nodes $\textbf{j}$ and $\textbf{l}$ in either direction and an exponential term that is a function of the regulation parameter $p$.
  With the homogeneous solution in Eq. \ref{eq:homo}, we can write the solution to the original non-homogeneous differential equation in Eq. \ref{eq:2ddgram}, that is, when $|\mathcal{D}| > 0$, and, noting that $W_{\textbf{i},\textbf{j}}(0) = 0$ for all $\textbf{i}$, $\textbf{j}$.
  \begin{equation}\label{eq:2dmultidriver}
    \begin{aligned}
      W_{\textbf{i},\textbf{j}}(t) &= \int_0^t e^{-2p\tau} \sum_{\textbf{k},\textbf{l}} I_{i_1-k_1}(2s\tau) I_{i_2-k_2}(2s\tau) I_{j_1-l_1}(2s\tau)\\
      &\times I_{j_2-l_2}(2s\tau) \sum_{\textbf{a}\in\mathcal{D}} \delta_{\textbf{k},\textbf{a}} \delta_{\textbf{l},\textbf{a}} d\tau \\
      &= \int_0^t e^{-2p\tau} \sum_{\textbf{a}\in \mathcal{D}} I_{i_1-a_1}(2s\tau) I_{i_2-a_2} (2s\tau)\\
      &\times I_{j_1-a_1}(2s\tau) I_{j_2-a_2}(2s\tau) d\tau
    \end{aligned}
  \end{equation}
  The MBFFKs which appear in the result of Eq. \ref{eq:2dmultidriver} have order equal to the distances between $i_1$, $j_1$ and $a_1$, as well as $i_2$, $j_2$ and $a_2$ for every node $\textbf{a} \in \mathcal{D}$.
  Finally, for the $2$-dimensional problem, we specialize the result in Eq. \ref{eq:2dmultidriver} to the case where $\mathcal{D} = \{(0,0)\}$, i.e., there is a single driver node located at the reference point of the entire lattice.
  \begin{equation}\label{eq:2donedriver}
    W_{\textbf{i},\textbf{j}}(t) = \int_0^t e^{-2p\tau} I_{i_1}(2s\tau) I_{i_2}(2s\tau) I_{j_1}(2s\tau) I_{j_2}(2s\tau) d\tau
  \end{equation}
  Plots of the integrand in Eq. \ref{eq:2donedriver} for several diagonal elements of $W_{\textbf{i},\textbf{i}}$ are shown in Fig. \ref{fig:cell}(b).
  When $p > 2ds$, i.e., the network is Hurwitz, the exponential term dominates the product of the four MBFFKs as, from property 2 of the MBFFK, they are strictly increasing functions of $\tau$.
  As for $p = 5$, the integrand in Eq. \ref{eq:2donedriver} decreases to zero exponentially in time, the integral expression for $W_{\textbf{i},\textbf{i}}$ converges to some finite value as seen in Fig. \ref{fig:cell}(c) for the same set of diagonal elements.\\
  \indent
  Consider the single driver single target problem as discussed above.
  The single driver is located at $\textbf{i}_{ref} = (0,0)$ and the single target is located at arbitrary node $\textbf{i}$.
  For this problem, the control energy is found to be \cite{klickstein2017energy},
  \begin{equation}
    E \propto \frac{1}{W_{\textbf{i},\textbf{i}}(t)} =  \left[ \int_0^t e^{-2p\tau} I_{i_1}^2(2s\tau) I_{i_2}^2 (2s\tau) d\tau \right]^{-1}
  \end{equation}
  As the target node moves further from the driver node, that is the order of the Bessel functions increase, from property 2 of the MBFFK, the integral decreases and so the energy increases (which can be visualized by examining the inverses of the curves in Fig. \ref{fig:cell}(c)).
  This increase of energy with distance between driver and target has been documented for general graphs \cite{wang2012optimizing}, but we show here the precise mechanism in lattice graphs.\\
  \indent
  An important consideration is the accuracy of the controllability Gramian for the infinite lattice when representing instead a finite lattice.
  In Fig. \ref{fig:error}(a), some elements of the controllability Gramian are shown with respect to their tetradic indices for a finite 2-dimensional lattice.
  The corresponding elements are shown in Fig.  \ref{fig:error}(b) computed using Eq. \ref{eq:2donedriver}.
  The absolute errors between these elements are shown in Fig. \ref{fig:error}(c) where the largest absolute error appears near the diagonal element corresponding to the driver node, $W_{0,0,0,0}(t)$.
  The relative error is shown in Fig. \ref{fig:error}(d), defined as the absolute error divided by the elements of the controllability Gramian of the finite graph, is largest for those elements of the controllability Gramian corresponding to those nodes closest to the edge of the finite lattice.
  Overall, if one is only targeting nodes not near the boundary of a finite lattice, computing the output controllability Gramian, $CW(t_f)C^T$, can be done efficiently using the exact solution given in Eq. \ref{eq:2dmultidriver}.
  \subsection{$d$-dimensional Regular Lattice}
  The results in the previous section for the $2$-dimensional regular lattice is extended to the more general $d$-dimensional regular lattice.
  A $d$-dimensional lattice exists in $\mathbb{Z}^d$ where a node exists at every integer coordinate and each node is indexed by the vector $\textbf{i} = (i_1,i_2,\ldots,i_d)$.
  Node $\textbf{i}$'s position in the lattice is denoted with respect to the reference node at $\textbf{i}_{ref} = (0,0,\ldots,0)$.
  Two nodes in the lattice are connected if they are distance one apart.\\
  \indent
  We extend our definition of the vector index incrementation and decrementation to accomodate nodes in a $d$-dimensional lattice.
  \begin{equation}
    \begin{aligned}
      \textbf{i}^{(k)} &= \left( \begin{array}{ccccc}
        i_1 & \cdots & i_k+1 & \cdots & i_d
      \end{array} \right), && 1 \leq k \leq d\\
      \textbf{i}^{(-k)} &= \left( \begin{array}{ccccc}
        i_1 & \cdots & i_k-1 & \cdots & i_d
      \end{array} \right), && 1 \leq k \leq d
    \end{aligned}
  \end{equation}
  The set of neighbors of node $\textbf{i}$ can be written efficiently as $\mathcal{N}_{\textbf{i}} = \{v_{\textbf{i}^{(k)}} | 1 \leq k \leq d\} \cup \{v_{\textbf{i}^{(-k)}} | 1 \leq k \leq d\}$.
  The differential equation that governs the evolution of the controllability Gramian for the infinite $d$-dimensional lattice graph is,
  \begin{equation}\label{eq:ddimdiff}
    \begin{aligned}
    \dot{W}_{\textbf{i},\textbf{j}}(t) &= -2pW_{\textbf{i},\textbf{j}} + s \sum_{k=1}^d \left( W_{\textbf{i}^{(k)},\textbf{j}} + W_{\textbf{i},\textbf{j}^{(k)}} + W_{\textbf{i}^{(-k)},\textbf{j}} + W_{\textbf{i},\textbf{j}^{(-k)}} \right)\\
    &+ \sum_{\textbf{k} \in \mathcal{D}} \delta_{\textbf{i},\textbf{k}} \delta_{\textbf{j},\textbf{k}}
    \end{aligned}
  \end{equation}
  The $2d$-dimensional DTFT is defined as,
  \begin{equation}
    \hat{W}_{\hat{\textbf{i}},\hat{\textbf{j}}}(t) = \sum_{\textbf{i},\textbf{j}} \prod_{k=1}^d e^{\mathcal{I} i_k \hat{i}_k} e^{\mathcal{I} j_k \hat{j}_k} W_{\textbf{i},\textbf{j}}(t)
  \end{equation}
  Applying the $2d$-dimensional DTFT to the differential equation for the controllability Gramian of the $d$-dimensional lattice graph in Eq. \ref{eq:ddimdiff} yields a result of similar form as for the $2$-dimensional lattice in Eq. \ref{eq:Whatij}.
  \begin{equation}
    \dot{\hat{W}}_{\hat{\textbf{i}},\hat{\textbf{j}}}(t) = \left(-2p + \sum_{k=1}^d \left(2s \cos \hat{i}_k + 2s \cos \hat{j}_k \right) \right) \hat{W}_{\hat{\textbf{i}},\hat{\textbf{j}}}(t)
  \end{equation}
  As this is a linear homogeneous equation for $\hat{W}_{\hat{\textbf{i}},\hat{\textbf{j}}}(t)$ decoupled from any other entry $\hat{W}_{\textbf{k},\textbf{l}}$ in the controllability Gramian we can solve for its evolution directly.
  \begin{equation}
    \hat{W}_{\hat{\textbf{i}},\hat{\textbf{j}}}(t) = e^{-2pt} \prod_{k=1}^d e^{2s\cos \hat{i}_k} e^{2s \cos \hat{j}_k} \hat{W}_{\hat{\textbf{i}},\hat{\textbf{j}}}(0)
  \end{equation}
  Finally, taking the inverse $2d$-dimensional DTFT to find $W_{\textbf{i},\textbf{j}}$,
  \begin{equation}\label{eq:ddimhomo}
    \begin{aligned}
      W_{\textbf{i},\textbf{j}}(t) &= \frac{1}{(2\pi)^{2d}} \int_{-\pi}^{\pi} \cdots \int_{-\pi}^{\pi} \prod_{k=1}^d e^{-\mathcal{I} i_k \hat{i}_k} e^{-\mathcal{I} j_k \hat{j}_k} e^{-2pt}\\
      &\times e^{2st \cos \hat{i}_k} e^{2st \cos \hat{j}_k} \sum_{\textbf{a},\textbf{b}} e^{\mathcal{I} \hat{i}_k a_k} e^{\mathcal{I} \hat{j}_k b_k} W_{\textbf{a},\textbf{b}}(0) d\hat{i}_k d\hat{j}_k\\
      &= \sum_{\textbf{a},\textbf{b}} W_{\textbf{a},\textbf{b}}(0) e^{-2pt} \prod_{k=1}^d \left( \frac{1}{2\pi} \int_{-\pi}^{\pi} e^{-\mathcal{I}(i_k-a_k) \hat{i}_k} e^{2st \cos \hat{i}_k} d\hat{i}_k \right)\\
      &\times \left( \frac{1}{2\pi} \int_{-\pi}^{\pi} e^{-\mathcal{I}(j_k b_k)\hat{j}_k} e^{2st \cos \hat{j}_k} d\hat{j}_k \right)\\
      &= \sum_{\textbf{a},\textbf{b}} e^{-2pt} \prod_{k=1}^d I_{i_k-a_k}(2st) I_{j_k-b_k}(2st) W_{\textbf{a},\textbf{b}}(0)
    \end{aligned}
  \end{equation}
  With the solution to the linear homogeneous equation (when $\mathcal{D} = \emptyset$ in Eq. \ref{eq:ddimdiff}), the solution to the nonhomogeneous differential equation (when $|\mathcal{D}| > 0$) is written as a convolution while also noting that $W_{\textbf{i},\textbf{j}}(0) = 0$ for all $\textbf{i},\textbf{j}$.
  \begin{equation}\label{eq:ddimmultidriver}
    \begin{aligned}
    W_{\textbf{i},\textbf{j}}(t) &= \int_0^t e^{-2p\tau} \sum_{\textbf{a},\textbf{b}} \prod_{k=1}^d I_{i_k-a_k}(2s\tau) I_{j_k-b_k}(2s\tau) \sum_{v_\textbf{r} \in \mathcal{D}} \delta_{\textbf{r},\textbf{a}} \delta_{\textbf{r},\textbf{b}}\\
    &= \int_0^t e^{-2p\tau} \sum_{v_\textbf{r} \in \mathcal{D}} \prod_{k=1}^d I_{i_k-r_k}(2s\tau) I_{j_k-r_k}(2s\tau) d\tau
    \end{aligned}
  \end{equation}
  The result in Eq. \ref{eq:ddimmultidriver} is analogous to Eq. \ref{eq:2dmultidriver} for the $d$-dimensional regular lattice, where each term in the summation is the offset of nodes $v_\textbf{i}$ and $v_\textbf{j}$ from the driver nodes $v_{\textbf{r}} \in \mathcal{D}$.
  \section{AN EXAMPLE}
  \subsection{Control Energy of a Finite Lattice}
  \begin{figure}
    \centering
    \includegraphics[width=\columnwidth]{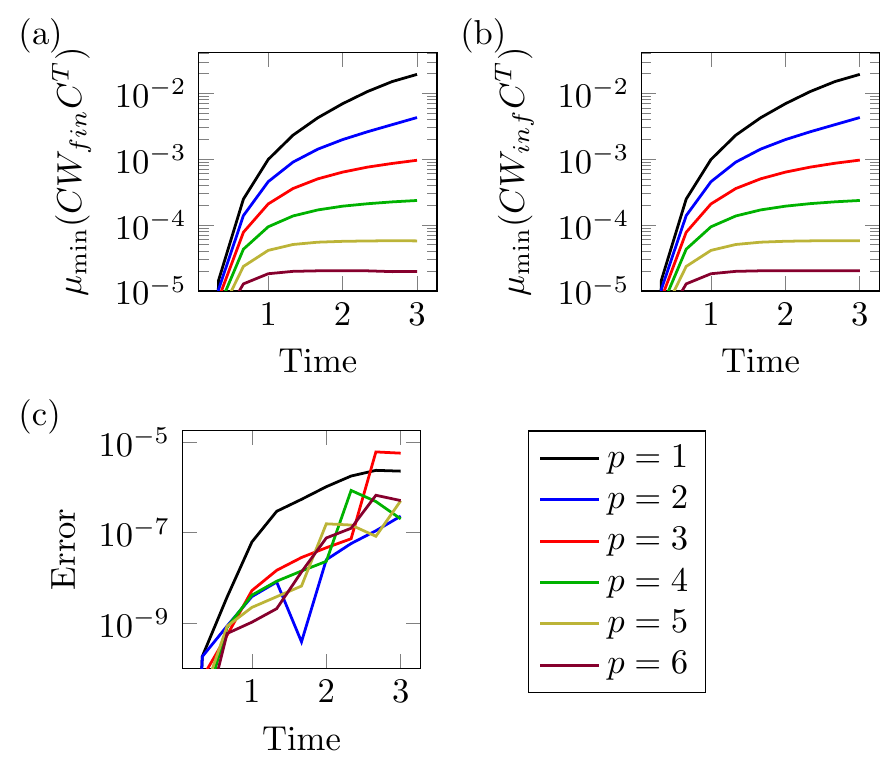}
    \caption{
      Comparison of the minimum eigenvalue of the output controllability Gramian for the analytic calculation (a) and the infinite calculation (b) for a variety of choices of time $t$ and regulation parameter $p$.
      For all simulations $s=1$.
      We choose the minimum eigenvalue as it plays a dominant role in the computation of the control energy for a general control action (which we call $\textbf{b}$ Eq. \ref{eq:energy}).
      The absolute error of the minimum eigenvalue when using the controllability Gramian of the infinite lattice (c) remains at least two orders of magnitude less than the value of the eigenvalue itself.
    }
    \label{fig:infinite}
  \end{figure}
  Consider a distributed system described by a 2-dimensional regular lattice of size $21 \times 21$ (so that the nodes at the edge of the lattice are distance at least distance ten from the central node).
  Each edge has weight $s = 1$, and self-loop magnitude $p$ which is varied.
  There is one driver node, $\mathcal{D} = \{(0,0)\}$, and three target nodes consisting of the driver node and two of its neighbors, $\mathcal{T} = \{(0,0),(1,0),(1,1)\}$.
  We compute both the controllability Gramian of the finite lattice using Eq. \ref{eq:fingram} and the controllability Gramian of the corresponding infinite lattice (with the same $p$, $s$, and $t_f$) using Eq. \ref{eq:2donedriver}.
  The smallest eigenvalue of the output controllability Gramian plays a dominant role in the expression for the control energy \cite{klickstein2017energy}.
  Let $\mu_i$ and $\textbf{z}_i$ be the $i$th eigenvalue and eigenvector of the output controllability Gramian $CW(t_f)C^T$, respectively.
  The control energy can be expressed in terms of the eigendecomposition of the output controllability Gramian.
  \begin{equation}
    E = \sum_{i=1}^{|\mathcal{T}|} \frac{1}{\mu_i} \left( \textbf{b}^T \textbf{z}_i \right)^2 \propto \frac{1}{\mu_{\min}}
  \end{equation}
  where $\mu_{\min}$ is the smallest eigenvalue that satisfies the inequalities $0 < \mu_{\min} \leq \mu_i$ as long as the triplet $(A,B,C)$ is output controllable which it is for our particlar case.\\
  \indent
  In Fig. \ref{fig:infinite}, we compare the minimum eigenvalue of the output controllability Gramian for both the finite lattice described above using Eq. \ref{eq:fingram} and the corresponding output controllability Gramian of the infinite 2-dimensional lattice using Eq. \ref{eq:2donedriver} in Fig. \ref{fig:infinite}.
  Both the finite lattice output controllability Gramian in Fig. \ref{fig:infinite}(a) and the infinite lattice output controllability Gramian have qualitatively similar minimum eigenvalues for various values of both time $t$ and of regulation parameter $p$.
  The difference between the minimum eigenvalue is shown in Fig. \ref{fig:infinite}(c) which we see overall is at least two orders of magnitude smaller.
  This suggests that the control energy can be reliably computed using the infinite 2-dimensional lattice approximation.\\
  \indent
  This approximation method is also computationally much more efficient.
  To compute the controllability Gramian for a network with $n$ nodes, we must simulate $\frac{n^2+n}{2}$ unique differential equations with Eq. \ref{eq:fingram} to compute the controllability Gramian (by exploiting its symmetry).
  For this moderate example with the finite lattice of a $21 \times 21$ lattice, which consists of $441$ nodes, we must simulate $97,461$ elements of the controllability Gramian.
  On the other hand, we only needed to compute 6 unique values to determine the output controllability Gramian to achieve a good approximation using Eq. \ref{eq:2donedriver}.
  When the target node set is small with respect to the number of nodes in the network, the type of approximation of the controllability Gramian presented here is not just important to understand the underlying mechanisms at work, but they also represent powerful time-saving numerical techniques.
  %
  %
  \section{CONCLUSION}
  We have derived the exact equation for the controllability Gramian for an infinite $d$-dimensional lattice with a finite set of driver nodes.
  From this result, we can compute the control energy when one wishes to drive a finite set of target nodes to some final state.
  We also demonstrated the application to finite lattice graphs as the relative error remains small away from the edges of the lattice.
  While there is an extensive literature on this subject \cite{shirin2017optimal,chen2016energy,ruths2014control} ours is one of the first analytical results in terms of computing the minimum control energy for large graphs.
  Overall, this paper provides a substantial step towards understanding the complex relationships between graph topology, distribution of driver nodes, selection of targets nodes, and the control energy needed to drive the target nodes to some final state.
  \section{ACKNOWLEDGMENTS}
  I. Klickstein would like to thank V. M. Kenkre for englightening conversations.
  This work is supported by the National Science Foundation through NSF grant CMMI-1400193, NSF grant CRISP-1541148, and ONR Award No. N00014-16-1-2637 as well as HDTRA1-13-1-0020.
  %
  %
  \bibliographystyle{IEEEtran}

\end{document}